\documentclass{article}
\pdfoutput=1
\usepackage{jheppub}
\usepackage{amsmath}
\usepackage{graphicx}
\usepackage{amssymb}
\usepackage{subfig}
\usepackage[utf8]{inputenc}
\usepackage{verbdef}

\title{Weakly Supervised Classification in High Energy Physics}

\author[a]{Lucio Mwinmaarong Dery}
\author[b]{Benjamin Nachman}
\author[c]{Francesco Rubbo}
\author[c]{Ariel Schwartzman}

\affiliation[a]{Physics Department, Stanford University, Stanford, CA, 94305, USA}
  
\affiliation[b]{Physics Division, Lawrence Berkeley National Laboratory, 1 Cyclotron Rd, Berkeley, CA, 94720, USA}  

\affiliation[c]{SLAC National Accelerator Laboratory, Stanford University, 2575 Sand Hill Rd, Menlo Park, CA, 94025, USA}

\emailAdd{ldery@slac.stanford.edu, bnachman@cern.ch, rubbo@slac.stanford.edu, sch@slac.stanford.edu}

\abstract{
As machine learning algorithms become increasingly sophisticated to exploit subtle features of the data, they often become more dependent on simulations.  This paper presents a new approach called {\it weakly supervised classification} in which class proportions are the only input into the machine learning algorithm.  Using one of the most challenging binary classification tasks in high energy physics - quark versus gluon tagging - we show that weakly supervised classification can match the performance of fully supervised algorithms.  Furthermore, by design, the new algorithm is insensitive to any mis-modeling of discriminating features in the data by the simulation.  Weakly supervised classification is a general procedure that can be applied to a wide variety of learning problems to boost performance and robustness when detailed simulations are not reliable or not available. 
}

\begin{document}
\maketitle
\flushbottom

\section{Introduction}

With the increasing complexity of theoretical models and computing power, many scientific projects increasingly rely on simulations to design analysis techniques.  This is especially true for high energy particle physics, where high fidelity Monte Carlo (MC) simulation is used to model physical processes at distances ranging from $10^{-25}$ meters all the way to the macroscopic dimensions of detectors.  However, just as models have become more complex, analysis techniques have also become more sophisticated.  Numerous, often subtle, features of events are combined using powerful supervised learning algorithms trained on large simulated (labeled) datasets. Despite these advances, there is no guarantee that techniques highly optimized in simulation are also optimal in nature.  One of the most ubiquitous analysis procedures is classifying events as originating from one of two different processes.  It is sometimes the case that one knows the proportions of each class better than the properties of each class that are useful for classification. {\it Weakly supervised classification} is a new machine learning paradigm for classification where training is performed directly on (unlabeled) data.

The task of training a classifier on multidimensional data based only on class proportions is highly under-constrained.  Neural network training is already a non-convex problem, but removing all local information about class labels significantly increases the difficulty of optimization.   However, the field of multi-instance learning (MIL)~\cite{DIETTERICH199731} has shown that local information is not necessarily needed for classification\footnote{See Ref.~\cite{Amores201381} for a review of recent work in MIL.}.   The setup of MIL is a series of sets (`bags') of individual instances without individual labels.  Consider the task of distinguishing two classes, called $A$ and $B$.  For the training set, it is known if a bag contains at least one instance of class $A$.  The algorithm is then optimized to identify the presence of at least one instance of class $A$ in an unseen bag.  Recent work has extended this procedure to identify the class of individual instances, still only training on bag-level labels~\cite{Kotzias:2015:GIL:2783258.2783380}. In this paper we make {\it supervision even weaker} in that bag-labels are only known on average.  In particular, all that is known to the training is the expected fraction of class $A$ in any particular bag. This paradigm is also referred to as Learning with Label Proportions (LLP)~\cite{NIPS2014_5453}.

High energy quarks and gluons produced in reactions at the Large Hadron Collider (LHC) result in collimated streams of particles traveling at nearly the speed of light, known as {\it jets}.
One of the most challenging classification tasks in high energy physics is to distinguish quark-induced jets from gluon-induced jets based on their radiation pattern.  There is an extensive literature exploring discriminating observables~\cite{Gallicchio:2011xq}. 
However, standard quark and gluon discriminants 
are known to be poorly modeled by state-of-the-art simulations~\cite{Aad:2014gea,Badger:2016bpw}.  Despite this, the fraction of quark and gluon jets in a given sample is often well-known.  At a fixed order in perturbation theory, the probability for an out-going parton to be a quark or a gluon depends on well-known parton distribution functions and matrix elements.  Therefore, quark versus gluon discrimination is well-suited for for weakly supervised classification and is therefore the main example used later in this paper.

This paper is organized as follows.
Section~\ref{sec:weaksupervision} formally introduces weakly supervised classification and describes how it is applied in practice.
The technique is illustrated using quark versus gluon jet tagging in Sec.~\ref{sec:qg}.
The paper ends in Sec.~\ref{sec:conclusions} with some concluding remarks.
The source code implemented to produce the results presented in this paper is available at \href{https://doi.org/10.5281/zenodo.322813}{DOI: 10.5281/zenodo.322813}.

\section{Weakly supervised classification}
\label{sec:weaksupervision}

Given a set of data originating from two classes labeled $0$ and $1$, 
the goal of classification is to construct a function $f:\mathbb{R}^n\rightarrow\{0,1\}$, 
where $n$ is the dimensionality of the feature space used to discriminate the two classes.  
In the traditional classification paradigm of fully supervised training, 
the function $f_\text{full}$ is built by minimizing a loss function like the following:

\begin{align}
\label{eq:full}
f_\text{full}=\text{argmin}_{f':\mathbb{R}^n\rightarrow\{0,1\}} \sum_{i=1}^N \ell\left(f'(x_i)-t_i\right),
\end{align}

\noindent where $N$ is the number of labeled data available for training, 
$\ell$ is a loss function with $\lim_{x\rightarrow0}\ell(x)=0$, and $t_i$ is the true label of example $i$.  
A common loss function is the squared error.  
In order to provide flexibility and stability, one often modifies the original problem to take 
$f:\mathbb{R}^n\rightarrow[0,1]$ and the output is interpreted as a probability for an event to be in class $0$ or $1$.  The ideal classifier that one tries to approximate with Eq.~\ref{eq:full} is based on the likelihood ratio $p(\vec{x}|0)/p(\vec{x}|1)$, where $p(\vec{x}|i)$ is the $n$-dimensional probability density for the feature vector $\vec{x}$ for the class $i\in\{0,1\}$.
{\it Weakly supervised classification} is a new paradigm in which instead of knowing the $t_i$, 
all that is known is the proportion of events in either class: $y=\sum_i t_i / N$.  
Thus, the weakly supervised $f_\text{weak}$ is given by

\begin{align}
\label{eq:weak}
f_\text{weak}=\text{argmin}_{f':\mathbb{R}^n\rightarrow [0,1]} \ell\left(\sum_{i=1}^N \frac{f'(x_i)}{N}-y\right).
\end{align}

\noindent The argument of Eq.~\ref{eq:weak} is non-convex, with many minima.  
In particular, the trivial solution $f'(x)=y$ results in a loss of zero.
However, using multiple batches of data with different proportions $y_k$ is sufficient to
collapse the solution space, so long as the distribution $p(\vec{x}|i;k)=p(\vec{x}|i)$, i.e. the distribution of the discriminating features for a particular class is the same in every batch $k$.  To build intuition for why there is any hope to solve this problem, consider a case where there are two batches $A$ and $B$ with proportions $y_A$ and $y_B$.  Consider an $n$-dimensional histogram where the $i^\text{th}$ dimension captures a discretized version of the $i^\text{th}$ discriminating feature.  If the $i^\text{th}$ dimension has $m_i$ bins, then the total number of bins in the histogram is $M=\sum_{i=1}^nm_i$.  One can always rearrange bins so that instead of an $n$-dimensional histogram with $m_i$ bins in the $i^\text{th}$ dimension, there is a one-dimensional histogram with $M$ bins.  As visualizing high dimensional histograms can be cumbersome, let $h_A$ be one-dimensional histograms with $M$ bins for the batch $A$ and $h_B$ be the corresponding histogram for batch $B$.  Then, for each bin $i$, one can write

\begin{align}
  \label{eq:system}
  h_{A,i} = y_Ah_{1,i} + (1-y_A) h_{0,i} \\
  h_{B,i} = y_Bh_{1,i} + (1-y_B) h_{0,i},
\end{align}

\noindent where $h_{X,i}$ is the content of the $i^\text{th}$ bin of the histogram $h_X$.  Except for contrived scenarios, Eq.~\ref{eq:system} will have a unique solution for $h_{0,i}$ and $h_{1,i}$, which are discretized versions of the probability densities  $p(\vec{x}|0)$ and $p(\vec{x}|1)$.  One can then form an (approximately) optimal classifier from the ratio of histograms with bin contents $h_{0,i}$/$h_{1,i}$.  If the number of dimensions is large, one can add a further step to use machine learning to approximate the optimal classifier from $h_{0,i}$ and $h_{1,i}$.  As a result, the problem is completely 	solvable.  Weakly supervised training combines the classification step with the first step and does so without binning.  Solving Eq.~\ref{eq:system} `by-hand' is intractable when $n$ is relatively large or the number of examples is relatively small.  It is also complicated when there are more than two batches (over-constrained).  These challenges are all naturally handled by the all-in-one machine learning approach of weakly supervised classification, as illustrated below.

In the weakly supervised training used in the following examples, $f'$ in Eq.~\ref{eq:weak} is parametrized as a three-layer neural network with three inputs, 
a hidden layer with 30 neurons, and a sigmoid output.
We use the Adam optimizer~\cite{DBLP:journals/corr/KingmaB14} in Keras~\cite{chollet2015keras} with a learning rate of 0.009 and train for 25 iterations.
As reference, we consider a traditional classifier 
\begin{align}
  \label{eq:fullqg}
  f_\text{full}=\text{argmin}_{f':\mathbb{R}^n\rightarrow [0,1]} \ell\left(f'(x_i)-t_i\right),
\end{align}
where $t_i$ labels the individual instances and $f'$ is
parametrized as a three-layer neural network with three inputs, a hidden layer with 10 neurons, 
and a sigmoid output.
Minimization is performed with stochastic gradient descent in Keras with a 
learning rate of 0.01 run for 40 iterations.
For each training, both networks are initialized with random weights, following a normal distribution.

\begin{figure}[h!]
\centering
\includegraphics[width=0.6\textwidth]{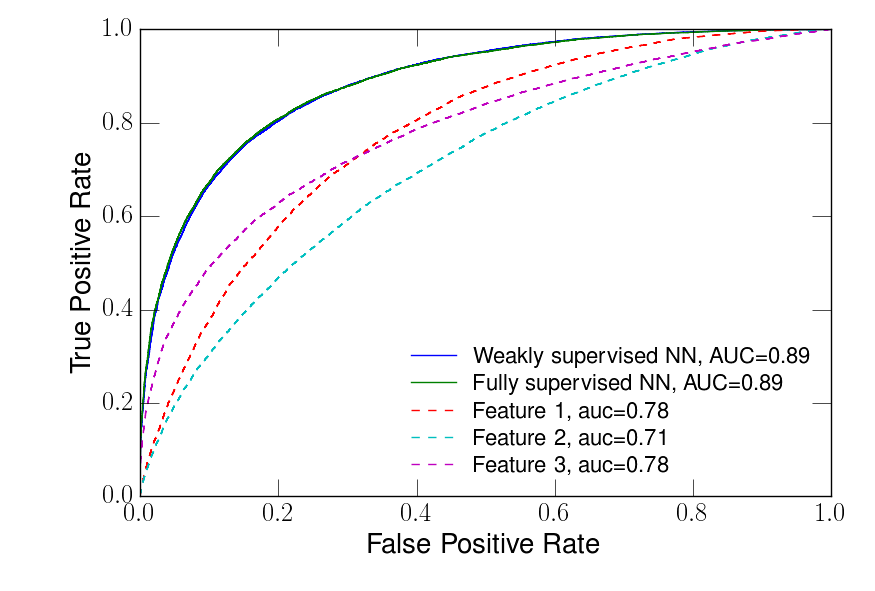}
\caption{Receiver Operator Characteristic (ROC) curves for instance classification, using three individual features and then combined using a fully supervised network and the weakly supervised classifier.  One performance metric is the Area Under the Curve (AUC) which is the integral of the ROC curve.}
\label{fig:rocs}
\end{figure}

\begin{table}[h!]
\begin{center}
\begin{tabular}{| c | c | c | c | c |}
\hline
Feature & $\mu_0$ & $\sigma_0$ & $\mu_1$ & $\sigma_1$ \\ 
 \hline
 \hline
 1 & 26 & 8 & 18 & 7 \\ 
 2 & 0.09 & 0.04 & 0.06 & 0.04 \\ 
 3 & 0.28 & 0.04 & 0.23 & 0.05 \\ 
 \hline
\end{tabular}
\end{center}
\caption{Mean ($\mu$) and standard deviation ($\sigma$) values of the normal distributions for class $0$ and $1$ of each feature.
}
\label{tab:musigma}
\end{table}

Figure~\ref{fig:rocs} shows the weakly supervised classifier performance when training with 9 subsets of data with proportions
between 0.2 and 0.4 compared with that of the fully supervised one.  Three features, labeled $1-3$ are constructed so that the distribution of feature $i$ given class $j$ follows a normal distribution with mean $\mu_{ij}$ and standard deviation $\sigma_{ij}$.  For reference, the values of $\mu_{ij}$ and $\sigma_{ij}$ used for the example shown in Fig.~\ref{fig:rocs} are in Table.~\ref{tab:musigma}.  Both the traditional and weakly supervised classifiers have the same Receiver Operator Characteristic (ROC) and thus have identical classification performance.  Note that the loss for weakly supervised classification is symmetric with respect to swapping the class assignment, therefore the classifier output for a given training can give higher values for class $0$, while
for a different training it would give higher value for class $1$.

\begin{figure}[h!]
\centering
\includegraphics[width=0.6\textwidth]{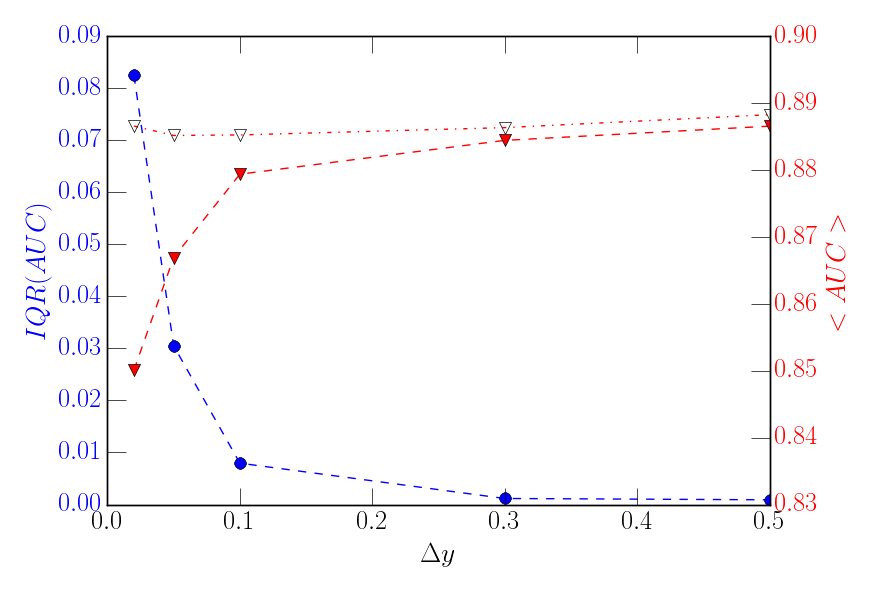}
\caption{Median (solid triangles) and interquartile range (solid dots) of the AUC as a function of the difference in proportions $\Delta y$
between the two subsets of the training sample. The proportion corrisponding to one subset is fixed to 0.2, while the other varies.
For each point the AUC is computed 100 times on the
same test set with different trainings, each performed with a random weight initialization. 
The maximum AUC for each point is also shown (hollow triangles).
}
\label{fig:stability}
\end{figure}

As with any machine learning algorithm with inherent randomness, the performance of a weakly supervised classifier has a stochastic component.  This is quantified by retraining the same network many times with different random number seeds in each iteration.  The interquartile range (IQR) over the Area Under the Curve (AUC) values for each training is a measure of the spread due to the inherent randomness.  Figure~\ref{fig:stability} shows the AUC IQR for the toy example with one proportion fixed to $0.2$ and the second proportion scanned from $0.2$ to $0.7$.  The stability improves as the difference between the class proportions increases.  In addition to the performance varying less as the proportions are further apart, the overall performance quantified by the median AUC (denoted by $\langle \text{AUC}\rangle$) also improves (increases).  The improvement in the median AUC is not as dramatic as the reduction in the AUC IQR, but it does suggest that it is (slightly) easier for the machine learning algorithm when the proportions are very different\footnote{Even when the proportions are within few percents,
stable performance can be achieved if multiple ($>2$) subsets with different proportions can be used for training.}.  This makes sense in the context of the two-step intuition-building paradigm given above: the algorithm can spend more attention on the classification task if it is easier to extract the class distributions.

\section{Example: quark and gluon jet discrimination}
\label{sec:qg}

Due to the strength of the strong force, there is a plethora of gluon jets produced at the LHC.  However, many processes result in mostly quark jets.  Prominent examples include the identification of hadronically decaying $W$ bosons~\cite{CMS:2014joa,Aad:2015owa}, jets associated with vector boson fusion~\cite{Khachatryan:2015bnx,Khachatryan:2014dea,Aaboud:2016cns}, and multi-quarks resulting from supersymmetry~\cite{Bhattacherjee:2016bpy}.  The references given here are the small number of public results that mention quark/gluon tagging, but there many more analyses that would benefit from a tagger if a robust technique existed.

The weakly supervised classification strategy is particularly useful for quark/gluon tagging because the fraction of quark jets for a particular set of events is well-known from parton distribution functions and matrix element calculations while useful discriminating features have not been computed to high accuracy and simulations often mis-model the data.  To illustrate this concrete example, quark and gluon jets are simulated and a weakly supervised classifier is trained on the generated event sample.  Unlike real data, in the simulated sample, we also know per-event labels which are used to additionally train a fully supervised classifier.  Events with $2\rightarrow 2$ quark-gluon scattering ({\it dijet} events) are simulated using the Pythia 8.18~\cite{Pythia8} event generator.  Jets are clustered using the anti-$k_t$ algorithm~\cite{Cacciari:2008gp} with distance parameter $R=0.4$ via the FastJet 3.1.3~\cite{fastjet} package.  Jets are classified as quark- or gluon-initiated by considering the type of the highest energy quark or gluon in the full generator event record that is inside a $0.3$ radius of the jet axis.   For simplicity, one transverse momentum range is considered: 45 GeV $<p_\text{T}<$ 55 GeV.  Additionally, there is a pseudo-rapidity requirement that mimics the usual detector acceptance for charged particle tracking: $|\eta|<2.1$.  Heuristically, gluons have twice as much strong-force charge as quark jets, resulting in more constituents and a broader radiation pattern.  Therefore, the following variables are useful for quark/gluon discrimination: the number of jet constituents $n$, the first radial moment in $p_\text{T}$ ({\it jet width}) $w$, and the fraction of the jet $p_\text{T}$ carried by the leading anti-$k_\text{T}$ $R=0.1$ subjet $f_0$.  The constituents $i$ considered for computing $n$ and $w$ are the hadrons in the jet with $p_\text{T}>500$ MeV.

\begin{figure}[h!]
\centering
\subfloat[][]{\includegraphics[width=0.475\textwidth]{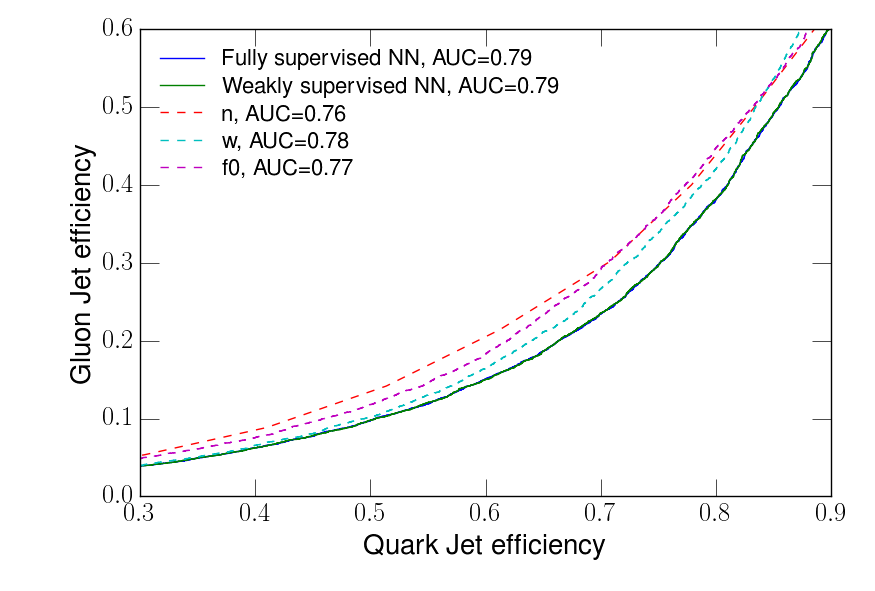}\label{rocs}}
\subfloat[][]{\includegraphics[width=0.475\textwidth]{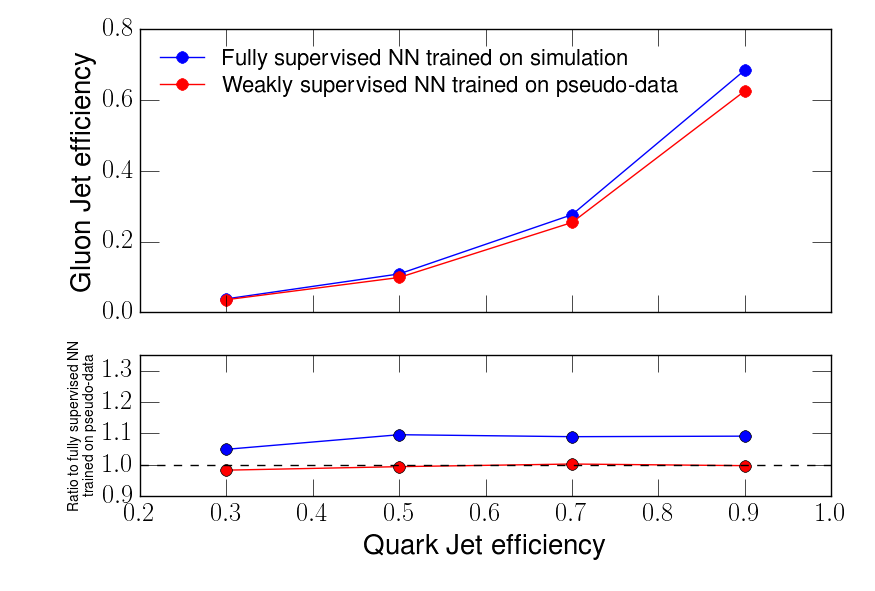}\label{rocs2}}
\caption{
Comparison of ROC curves for quark/gluon jet discrimination 
using a fully supervised classifier or a weakly supervised classifier.
In \protect\subref{rocs} the fully and weakly supervised classifiers are trained on identical simulated data and evaluated on a test sample drawn from the same population.
The weakly supervised classifier matches the performance of the fully supervised one.
The curves corresponding to the three input observables used as discriminant are shown as reference.
In \protect\subref{rocs2}, the fully supervised classifier (blue line) is trained on a labeled simulated training sample.
The weakly supervised classifier (red line) is trained on an unlabeled pseudo-data training sample.
In both cases, the performance is evaluated on the same pseudo-data test sample.
The ratios to the performance of a fully supervised classifier trained on a labeled pseudo-data sample are shown in the bottom pad.
}
\label{fig:rocs_gq}
\end{figure}

A weakly supervised classifier with one hidden layer of size 30 is trained by considering 12 bins of the
distribution of the absolute difference in pseudorapidity between the two jets~\cite{Aad:2016oit}.  The proportion of
quark initiated jets varies between 0.21 and 0.32.  Figure \ref{fig:rocs_gq} shows that, while the individual observables perform differently in the high
or low gluon efficiency (true positive rate) regimes, their combination in a NN gives consistently better performance.
The weakly supervised classifier matches the performance of the fully supervised NN, despite only knowing sample proportions instead of individual event labels.  
By construction the weakly supervised classifier is also robust against a realistic amount of mis-modeling in the input variables.  
This feature is tested by building a pseudo-data sample where the probability distributions of $n$ and $w$ are distorted in the training sample to emulate the difference in efficiency measured in Ref.~\cite{Aad:2014gea}.  The study in Ref.~\cite{Aad:2014gea} found that a classifier extracted from simulation is more powerful than one extracted from the data. This is reflected in the results presented in the right plot of Fig.~\ref{fig:rocs_gq}.  When a fully supervised classifier is trained on a sample generated with the same distribution as the test sample (mimicking training and testing on simulation), it achieves a better performance than when trained on the original sample and tested on the distorted pseudo-data (mimicking training on simulation and testing on data).  In contrast, the weakly supervised classifier can be trained directly on the distorted pseudo-data sample (representing the data) so is insensitive to the mismodeling of the input variables.  This results in a 10\% bias from the standard procedure that is avoided by the weakly supervised classifier.  Even larger differences may be expected from this and other classification tasks that utilize even more input features or are more mis-modeled.  The weakly supervised classifier is robust and outperforms the standard supervised learning trained on simulation.   

\begin{figure}[h!]
\centering
\includegraphics[width=0.6\textwidth]{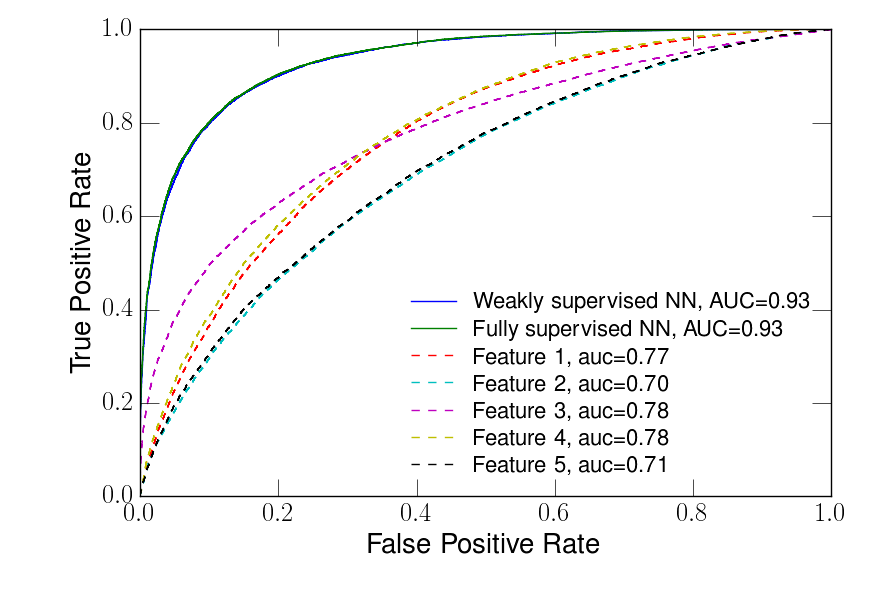}
\caption{ROC curves for instance classification using five individual features and then combined using a fully supervised network and the weakly supervised classifier.  
}
\label{fig:rocs_many}
\end{figure}

\section{Conclusions}
\label{sec:conclusions}

We have presented a new approach to classification with NN in cases where class proportions are known but individual labels are not readily available.   This weakly supervised classification has broad applicability and has been demonstrated in one important discrimination task in high energy physics: quark versus gluon jet tagging.  In the quark/gluon and related contexts, weakly supervised classification provides a robust and powerful approach because it can be directly trained on examples from (unlabeled) data instead of (labeled, but unreliable) simulation.  The examples presented so far have used a small number of input features to illustrate the ideas, but there is no algorithmic limitation on the number of features.  Figure~\ref{fig:rocs_many} is a simple extension of Fig.~\ref{fig:rocs} with 5 features instead of 3; in future work, we will study the extension to many more features (tens to hundreds).  This paper has laid the conceptual groundwork for this new tool that has started a new classification paradigm that can be applied to a wide variety of learning problems to boost performance and robustness when detailed simulations are not reliable or not available. 
 
\section{Acknowledgments}

This work is supported by the Stanford Data Science Initiative and by the US Department of Energy (DOE) under grant DE-AC02-76SF00515.
We would like to thank Russell Stewart for useful discussions about label-free supervision strategies and non-convex optimization problems and Gilles Louppe for useful discussion about related work on learning from label proportions.

\bibliographystyle{JHEP-2}
\bibliography{myrefs}

\end{document}